# Phase fluctuations in a strongly disordered s-wave NbN superconductor close to the metal-insulator transition

Mintu Mondal[1], Anand Kamlapure[1*], Madhavi Chand[1†], Garima Saraswat[1], Sanjeev Kumar[1], John Jesudasan[1], L. Benfatto[2], Vikram Tripathi[1] and Pratap Raychaudhuri[1‡]

[1]*Tata Institute of Fundamental Research, Homi Bhabha Rd., Colaba, Mumbai 400005, India.*
[2] *ISC-CNR and Department of Physics, University of Rome "La Sapienza", P.le A. Moro 5, 00185 Rome, Italy.*

We explore the role of phase fluctuations in a 3-dimensional *s*-wave superconductor, NbN, as we approach the critical disorder for the destruction of the superconducting state. Scanning tunneling spectroscopy measurements reveal that in the presence of strong disorder, a finite gap in the electronic spectrum continues to persist at temperatures well above $T_c$. In the same range of disorder, the superfluid density is strongly suppressed at low temperatures and evolves towards a linear-T variation at higher temperatures. These observations provide strong evidence that phase fluctuations play a central role in the formation of a pseudogap state in a disordered conventional superconductor.

---

* ask@tifr.res.in
† chand@tifr.res.in
‡ pratap@tifr.res.in

Understanding the role of interaction and disorder is at the heart of understanding many-body quantum systems such as high-temperature superconductors, quantum Hall effect and superfluid helium. Over the past few decades, it has been understood that the ground state in conventional superconductors, once thought to be resilient to disorder[1], gets destabilized at a critical value of disorder, giving rise to unusual metallic and insulating states[2,3]. However, several recent experiments indicate that even after the global superconductivity is destroyed, the material continues to possess some of the fundamental properties commonly associated with the superconducting state. These include the dramatic observation of magnetic flux quantization in disorder driven insulating Bi films[4], finite high-frequency superfluid stiffness above $T_c$ in amorphous $InO_x$ films[5] and the formation of a "pseudogapped" state[6] above $T_c$ characterized by a finite gap in the electronic spectrum but no global superconductivity. Recent theoretical investigations[7] also indicate that strong superconducting correlation can persist even after global superconductivity is destroyed.

The superconducting state is characterized by a complex order parameter of the form, $\Psi = |\Delta|e^{i\phi}$. There are three kinds of excitations over the superconducting ground state that can ultimately destroy the superconducting order. (i) Quasiparticle excitations which primarily affect the amplitude of the order parameter; (ii) quantum phase fluctuations (QPF) associated with the number-phase uncertainty relation, according to which phase coherence between neighboring regions implies large number fluctuations and hence large Coulomb energies, and (iii) classical phase fluctuations (CPF) between adjacent regions in the superconductor caused by thermal excitations. In conventional superconductors in the "clean" limit, phase fluctuations[8] play a negligible role and it is sufficient to consider quasiparticle excitations alone, which are well described by Bardeen-Cooper-Shrieffer (BCS) and Eliashberg mean-field theories[9]. However, strongly disordered superconductors are characterized by poor screening of the Coulomb interactions[10] and small superfluid density due to disorder scattering[9], both of which render the superconductor more susceptible to phase fluctuations. Since strong phase fluctuations can destroy the global superconducting order well before the amplitude of the order parameter goes to zero, a larger role of phase fluctuations in strongly disordered superconductors could explain the seemingly contradictory observations of apparent superconducting properties persisting in the normal state of the material. This scenario also finds support from numerical simulations[11] which indicate that the superconducting state can get destroyed by strong phase fluctuations between domains that spontaneously form in a superconductor in the presence of strong disorder.

In this letter, we explore the role of phase fluctuations on the superconducting properties of strongly disordered epitaxial NbN films, using a combination of scanning tunneling spectroscopy (STS) and penetration depth ($\lambda$) measurements. The effective disorder, characterized by $k_Fl$ (where $k_F$ is the Fermi wave-vector and $l$ is the electronic mean free path), ranges from $k_Fl$~1.24-10.12. Within this range, $T_c$ (defined as the temperature where the resistance reaches 1% of its normal state value) decreases from 16.8K to <300mK. In films with $T_c \lesssim 6K$, STS measurements show that the dip in the tunneling density of states (DOS) associated with the superconducting energy gap, persists at temperatures well above $T_c$. In the same range of disorder, penetration depth measurements reveal suppression of superfluid density ($n_s$) compared to a BCS estimate, and a gradual evolution towards linear temperature variation of $n_s$. These observations provide compelling evidence of the role of phase fluctuations in strongly disordered s-wave superconductors.

Our samples consist of 3D epitaxial NbN films with thickness ≥ 50nm grown through reactive magnetron sputtering on single crystalline MgO substrates. The thickness of all the films is much larger than the dirty limit coherence length ($\xi \sim 4\text{-}8nm$)[12]. The disorder in these films can be tuned over a large range by controlling the deposition conditions, without destroying the crystalline structure of NbN and the epitaxial nature of the films[13,14]. Microscopically, the disorder in our films results from Nb vacancies in the crystalline NbN lattice, which is controlled by controlling the Nb/N ratio in the plasma[13,14]. For each film, $k_Fl$ is determined from the resistivity ($\rho$) and Hall coefficient ($R_H$) measured at 285K using the free electron formula $k_Fl = \{(3\pi^2)^{2/3}\hbar[R_H(285K)]^{1/3}\}/[\rho(285K)e^{5/3}]$ ($\hbar$ is the Plank's constant and $e$ is the electronic charge). The magnetic penetration depth ($\lambda$) is measured using a low-frequency (60kHz) two coil mutual inductance technique[15] on 8mm diameter circular films patterned using a shadow mask. This method allows measurement of the absolute value of $\lambda$ over the entire temperature range without any prior assumption regarding the temperature dependence of $\lambda$. The temperature dependence of the tunneling DOS is probed from the measurement of tunneling conductance ($G(V) = \frac{dI}{dV}\big|_V$) between the superconductor and a normal metal (Pt-Ir) tip using a high-vacuum low-temperature scanning tunneling microscope[16] (STM) operating down to 2.6K. Samples used for STM measurement are grown in-situ in a sputtering chamber connected to the STM. Tunneling measurements were also performed on disordered NbN/insulator/Ag planar tunnel junctions[17] at 500mK in a conventional $^3$He cryostat.

Fig. 1(a) shows the conductivity ($\sigma$) as a function of temperature for a series of NbN films. While most of the films display an unusual temperature variation of $\sigma$ in the normal state[14] $\left(\frac{d\sigma}{dT} > 0\right)$, even in the most disordered film (inset 1(a)), the variation in $\sigma$-$T$ (after superconductivity is suppressed by the application of a magnetic field) clearly suggests that $\sigma(T\rightarrow 0)\neq 0$. Therefore, we infer that all our samples are on the metallic side of the Anderson metal-insulator transition (MIT). We observed a remarkable correlation between $T_c$, the normal state conductivity just before the onset of the superconducting transition ($\sigma_0$) and $k_Fl$ across all our films grown over a period of several months. Fig. 1(b) shows that both $\sigma_0$ and $T_c$ asymptotically go to zero in the limit $k_Fl\rightarrow 1$. Therefore, within the accuracy of our measurements, the critical disorder for the destruction of superconductivity coincides with the Anderson metal-insulator transition (MIT). It is interesting to note that while a direct transition from a superconducting to an insulating state with disorder is ubiquitous in ultrathin 2-dimensional films[2], in 3-D films, both superconductor-normal metal and superconductor-insulator transitions have been observed and theoretically predicted[18,10,19]. At present, we do not know whether this coincidence is merely accidental or has a deeper significance. Experimentally, a similar scenario is reported for 3-D boron-doped diamond epitaxial films[20] where the MIT coincides with superconductor-insulator transition at the same level of boron doping.

To understand the temperature evolution of the tunneling DOS in the presence of strong disorder, STS measurements were performed on several samples with different disorder. Figures 2(a)-(d) show the temperature dependence of the tunneling DOS in the form of an intensity plot of the normalized conductance, $G(V)/G_N$ (where $G_N$ is the conductivity at high bias), as a function of temperature and bias voltage for 4 different samples. For each temperature and bias, the tunneling conductance is spatially averaged over 32 points taken at equal intervals along a 150nm line on the surface. The resistance vs. temperature for the same samples measured after taking them out from the STM are shown in the same panels. At the lowest temperature, the tunneling conductance of the first 3 samples (shown in *inset*) show a dip close to zero bias and two symmetric peaks, as expected from BCS theory. For the sample with $T_c$~11.9K which is far from the critical disorder, a flat metallic density of states (DOS) is restored just above $T_c$. However, with increase in disorder a "pseudogapped" phase emerges, where the dip in the DOS persists above $T_c$ and metallic DOS is restored at a temperature much higher than $T_c$. In the most disordered of these 4 samples with $T_c$~1.65K, the pseudogap persists well beyond the onset of the superconducting transition, up to ~6.5K. In this context we would like to note that even above the pseudogap temperature, in all our samples, a "V" shape in the conductance spectra extends up to bias voltages much higher than the characteristic superconducting energy gap[21] ($\Delta$). This feature, which is not associated with superconductivity, arises from Altshuler-Aronov (AA) type electron-electron interactions in the normal state. In order to extract the pseudogap feature associated with superconductivity in the most disordered sample ($T_c$~1.65K), we subtract the AA background using the spectra obtained at 8K (which is above the pseudogap temperature), from the spectra obtained at 2.65K (above the onset of the superconducting transition). The resulting curve $G_{sub}(V)\left(=\left.\frac{G(V)}{G_N}\right|_{T=2.65K}-\left.\frac{G(V)}{G_N}\right|_{T=8K}+0.76\right)$ (right inset of Fig. 2(d)) clearly reveals the presence of broadened coherence peaks around 2 mV confirming the superconducting origin of the pseudogap feature. The persistence of the superconducting energy gap and coherence peaks at temperatures well above $T_c$ clearly indicates that strong superconducting correlations persist even after the resistive state has been restored. In addition, the individual line scans[22] obtained at the lowest temperatures, reveal that the superconducting state becomes progressively inhomogeneous as the disorder is increased[11].

To investigate whether the "pseudogap" behavior is related to strong phase fluctuations we now focus on the penetration depth measurements. $n_s$ which is related to $\lambda$ through the London relation $n_s = m^*/(\mu_0 e^2 \lambda^2)$ ($m^*$ is the effective mass of the electron), gets strongly renormalized[23,24] in the presence of phase fluctuations. Figs. 3(a) and 3(b) show the temperature variation of $\lambda^{-2}$(T) in a series of NbN films with different $T_c$. For the films with low disorder the observed temperature variation agrees well with the dirty-limit BCS expression[9],

$$\frac{\lambda^{-2}(T)}{\lambda^{-2}(0)} = \frac{\Delta(T)}{\Delta(0)}\tanh\left(\frac{\Delta(T)}{2k_B T}\right),$$ where $k_B$ is the Boltzman constant. However, as the disorder increases two effects become noticeable: (i) $\lambda^{-2}$ in the limit $T\rightarrow 0$, decreases rapidly (by two orders of magnitude in the range $T_c\sim$15.8K to 2.27K) and (ii) $\lambda^{-2}(T)$ decreases faster that the expected BCS temperature variation. First we concentrate on the magnitude of $\lambda^{-2}(0)$. In the absence of phase fluctuations, $\lambda^{-2}(0)$ is reduced by disorder scattering according to the BCS relation[9], $\lambda^{-2}(0)_{BCS} = \frac{\pi\mu_0\Delta(0)\sigma_0}{\hbar}$. For NbN, we find $\Delta(0)\approx 2.05 k_B T_c$ from tunneling measurements performed at low temperatures ($T<0.2T_c$) on planar tunnel junctions fabricated on a large number of samples with different disorder[25] (inset Fig. 3(c)). Fig. 3(c) shows that $\lambda^{-2}(0)\approx\lambda^{-2}(0)_{BCS}$ within experimental error for samples with $T_c>6$K. However, as we approach the critical disorder $\lambda^{-2}(0)$ becomes gradually smaller than $\lambda^{-2}(0)_{BCS}$, reaching a value which is 50% of the BCS estimate for the most disordered sample of this series ($T_c\sim$2.27K). Focusing on the temperature dependence, we observe that while in all samples $\lambda^{-2}$(T) saturates as $T\rightarrow 0$, with increase in temperature, it shows a gradual evolution towards a linear-T variation for samples with $T_c<6$K. This linear variation is most clearly observed in the sample with $T_c\sim$2.27K, shown in Fig. 3(d).

Since the suppression of $\lambda(0)^{-2}$ from its BCS value and linear dependence of $\lambda^{-2}$ (T) with temperature are characteristic features associated with QPF and CPF[26] respectively, we now try to assess whether phase fluctuations can quantitatively account for the observed behavior. The importance of quantum and classical phase fluctuations is determined by two energy scales[8]: The Coulomb energy $E_c$, and the superfluid stiffness, $J$ ($\propto n_s$). For a 3-D superconductor these two quantities can be estimated from the relations[24],

$$E_c = \frac{16\pi e^2}{\varepsilon_\infty a} \text{ and } J = \frac{\hbar^2 a n_s}{4m^*}, \qquad (1)$$

where $\varepsilon_\infty$ is the background dielectric constant from the lattice and $a$ is the characteristic length scale for phase fluctuations. The suppression of $\lambda^{-2}(0)$ from QPF over its bare value can be estimated using the self consistent harmonic approximation[27] which predicts (in 3-D), $n_s(T=0)/n_{s0}(T=0) = e^{-\Delta\theta^2(T=0)/6}$ ($n_{s0}$ is the bare value in the absence of phase fluctuations), where $\Delta\theta^2(T=0) = \frac{1}{2}\sqrt{\frac{E_c}{J}}$. At the same time, the energy scale above which phase fluctuations become classical is given by the Josephson plasma frequency $\left(\hbar\omega_p = \sqrt{4\pi e^2 n_s/m^*\varepsilon_\infty} = \sqrt{E_c J}\right)$. We now calculate these values for the sample with $T_c \sim 2.27$K and compare with the data. For NbN, we estimate $\varepsilon_\infty \approx 30$ from the plasma frequency (12600 cm$^{-1}$) measured[28] at low temperatures. Taking the characteristic phase fluctuation length-scale $a \approx \xi \sim 8$nm [ref.12], we obtain, $E_c \approx 0.3$eV and $J \approx 0.14$meV at T=0. This corresponds to $n_s(T=0)/n_{s0}(T=0) \approx 0.02$. While this numerical value is likely to have some inaccuracy due to the exponential amplification of any error in our estimate of $E_c$ or $J$, the important point to note is that this suppression is much larger than our experimental estimate, $\lambda^{-2}(0)/\lambda^{-2}_{BCS}(0) \approx 0.5$. On the other hand $\hbar\omega_p \sim 6.5$ meV$\equiv 75$ K, so that one should conclude that CPF cannot be responsible for the observed linear temperature dependence of $\lambda^{-2}(T)$ in this sample.

These two apparent contradictions can be resolved by considering the role of dissipation. In *d*-wave superconductors, the presence of low energy dissipation has been theoretically predicted[29] and experimentally observed from microwave[30] and terahertz conductivity[31] measurements. Recent microwave measurements[32] on amorphous $InO_x$ films reveal that the conductivity remains finite at low frequencies even below $T_c$, implying that low energy dissipation can also be present in strongly disordered *s*-wave superconductors[33]. While the origin of this dissipation is not clear at present, the presence of dissipation has several effects on phase fluctuations[24]: (i) QPF are less effective in suppressing $n_s$; (ii) QPF contribute to a $T^2$ temperature depletion of $n_s$ of the form $n_s/n_{s0} = 1 - BT^2$ at low temperature where B is directly proportional to the dissipation and (iii) the crossover to the usual linear temperature dependence of $n_s$ due to CPF, $n_s/n_{s0} = 1 - (T/6J)$, occurs above a characteristic temperature that is much smaller than $\hbar\omega_p/k_B$, and scales approximately as $T_{cl} \approx J/\bar{\sigma}$, where $\bar{\sigma}$ is a dimensionless measure of the residual conductivity in the SC state. In the sample with $T_c$~2.27K, the $T^2$ variation of $\lambda(T)^{-2}/\lambda(0)^{-2}$ can be clearly resolved below 650mK (*inset* Fig. 3(d)). In the same sample, the slope of the linear-T region is 3 times larger than the slope estimated from the value of *J* calculated for T=0. This discrepancy is however not very serious considering the approximations involved. It is also important to note that at finite temperatures $n_{s0}$ itself will get renormalized due to quasiparticle excitations. With decrease in disorder, the effect of quasiparticle excitations will eventually dominate over the phase fluctuation corrections, thereby recovering the usual BCS temperature dependence in the low disorder limit, consistent with the data presented in Fig. 3(a). Since thermal phase fluctuations eventually lead to the destruction of the superconducting state at a temperature less than the mean field transition temperature, the increased role of phase fluctuation could naturally explain the observation of a pseudogapped state in strongly disordered NbN films. We would also like to note that in all the disordered samples $\lambda^{-2}(T)$ shows an abrupt downturn close to $T_c$. While such a downturn in ultrathin superconducting films is associated with the Kosterlitz-Thouless (KT) transition – smeared by disorder[15] – our samples are well in the 3D regime where a KT transition is not expected. At present we do not know the origin of this behavior.

To summarize, we observe a progressive increase in phase fluctuations and the formation of a pseudogap state in disordered NbN thin films as the disorder is increased towards the Anderson limit. The observation of strong phase fluctuations below $T_c$ in the same range of disorder where we observe the emergence of a pronounced pseudogap phase above $T_c$ supports the scenario where superconducting transition is governed by a phase disordering transition. The remaining question is on the role of amplitude fluctuations arising from non-equilibrium Cooper pairs, which can also produce a dip in the density of states in the tunneling DOS above $T_c$. While our experiments cannot rule out the possibility of amplitude fluctuations also playing a role (as suggested in a recent theoretical work[34]), within a standard Ginzburg-Landau (G-L) approach this effect is very small[35] and cannot account for the pronounced pseudogap feature observed in our samples. However, it would be worthwhile to critically explore the applicability of conventional G-L theory close to the MIT to assess the role of amplitude fluctuations.

In conclusion, we present experimental evidence of strong phase fluctuations and the formation of a pseudogapped state above $T_c$ in strongly disordered NbN thin films. These results provide new insight into several recent experiments[4,5,6] where strong superconducting correlations are seen to persist in the disorder driven metallic and insulating state. It would also be interesting to use such a comparative analysis between tunneling and penetration-depth experiments in underdoped high-$T_c$ cuprates, to understand the extent to which the "pseudogap" state in these materials can be understood within a phase-fluctuations scenario.

We would like to thank T. V. Ramakrishnan, S. Mandal and M. V. Feigelman for enlightening discussions and V. Bagwe and S. P. Pai for technical help.

[32] R. W. Crane et al., Phys. Rev. B **75,** 094506 (2007).

[33] Indirect evidence of the presence of low energy dissipation in our strongly disordered films is obtained from the tunneling DOS measured at low temperatures; for details see EPAPS, section 2.

[34] J. L. Tallon, J. G. Storey and J. W. Loram, arXiv:0908.4428v1.

[35] In the dirty-limit, the dip in G(V) at V=0 is expected to be of the order of $(\delta G/G)_{V=0} = \sqrt{Gi} \ln[(T-T_c)/T_c]$, where $Gi$ is the Ginzburg number given by, $Gi = (1.6/(k_F l)^3)(T_c/E_F)$; A. Larkin and A. Varlamov, *Theory of Fluctuations in Superconductors,* (Clarendon Press, Oxford, 2005).

*Figure captions:*

**Figure 1.** (a) $\sigma$ vs. $T$ for films with $k_Fl \sim$ 10.12, 8.82, 8.13, 8.01, 5.5, 4.98, 3.65, 3.27, 2.21, 1.68, 1.58 and 1.24. The *inset* shows an expanded view of $\sigma(T)$ for the film with $k_Fl \sim 1.24$ in zero field and in a magnetic field of 5T. (b) $\sigma_0$ and $T_c$ as a function of $k_Fl$.

**Figure 2. (a-d)** Intensity plots of the normalized tunneling conductance $(G(V)/G_N)$ as a function of temperature and applied bias voltage for four different films (upper panels), along with the temperature variation of resistance (*R-T*) in the same temperature range (lower panels). The vertical dotted line in the upper panel corresponds to $T_c$ measured from R-T. The insets of panels (a-c) and left inset of panel (d) show representative tunneling spectra in the superconducting state (blue), in the pseudogap region (green) and at the highest temperature of our measurements (red). The right inset of panel (d) shows an expanded view of R-T close to the superconducting transition.

**Figure 3.** (a)-(b) Temperature variation in $\lambda^{-2}(T)$ for NbN films with different disorder; the solid black lines are the expected temperature variation from dirty limit BCS theory. (c) $\lambda^{-2}(0)$ and $\lambda^{-2}(0)_{BCS}$ as function of $T_c$; the *inset* shows the variation $\Delta(0)$ as function of $T_c$. (d) Temperature variation of $(\lambda(T)/\lambda(0))^{-2}$ for the NbN film with $T_c = 2.27$K; the green solid line is a fit to the $T^2$ dependence of $\lambda^{-2}(T)$ at low temperature (T≤0.65K) and the red line shows the T dependence at higher temperature; the *inset* shows an expanded view of the $T^2$ dependence at low temperatures.

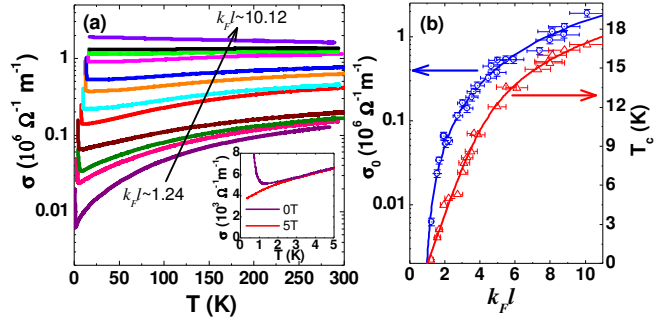

**Figure 1.** (a) $\sigma$ vs. $T$ for films with $k_Fl \sim$ 10.12, 8.82, 8.13, 8.01, 5.5, 4.98, 3.65, 3.27, 2.21, 1.68, 1.58 and 1.24. The *inset* shows an expanded view of σ(T) for the film with $k_Fl\sim$1.24 in zero field and in a magnetic field of 5T. (b) $\sigma_0$ and $T_c$ as a function of $k_Fl$.

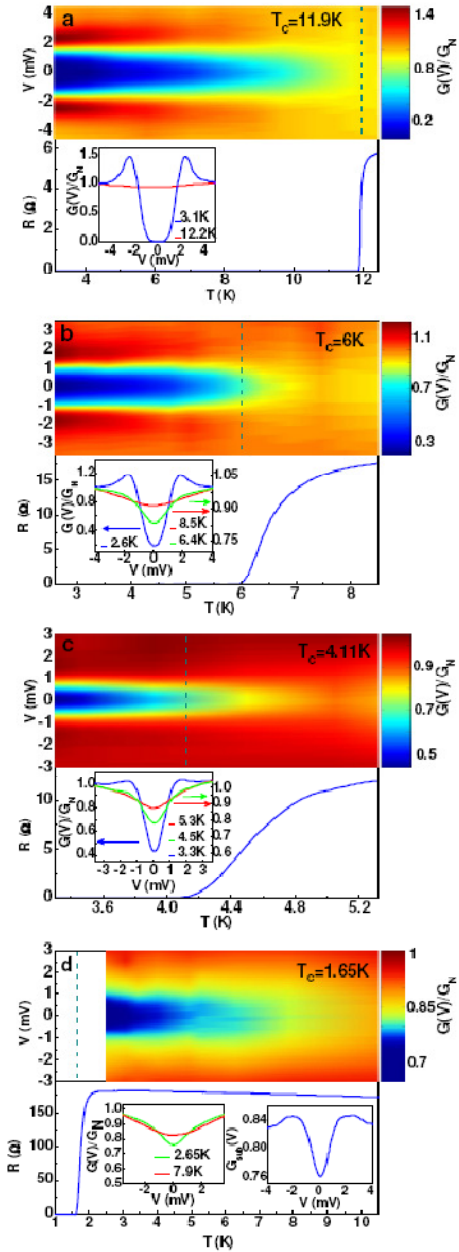

**Figure 2. (a-d)** Intensity plots of the normalized tunneling conductance (G(V)/G$_N$) as a function of temperature and applied bias voltage for four different films (upper panels), along with the temperature variation of resistance (*R-T*) in the same temperature range (lower panels). The vertical dotted line in the upper panel corresponds to $T_c$ measured from R-T. The insets of panels (a-c) and left inset of panel (d) show representative tunneling spectra in the superconducting state (blue), in the pseudogap state (green) and above the pseudogap temperature (red). The right inset of panel (d) the spectra at 2.65K after subtracting the Altshuler-Aronov background.

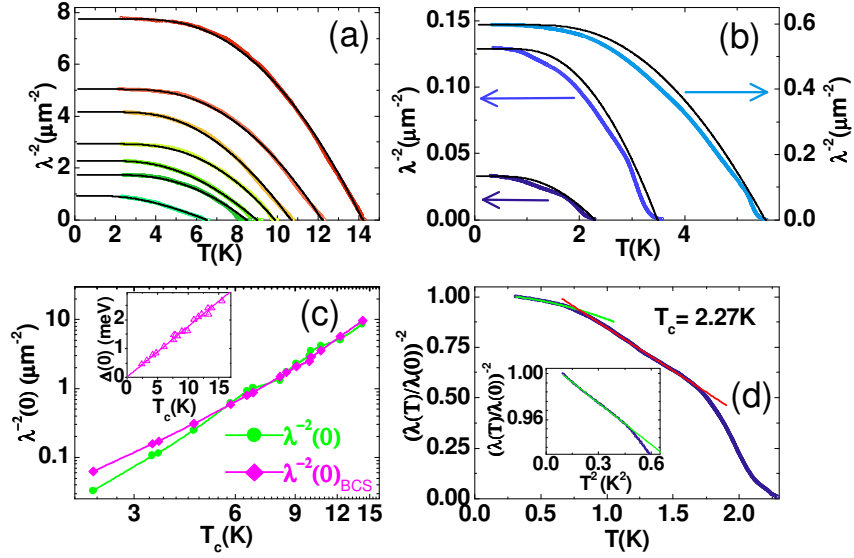

**Figure 3.** (a)-(b) Temperature variation in $\lambda^{-2}(T)$ for NbN films with different disorder; the solid black lines are the expected temperature variation from dirty limit BCS theory. (c) $\lambda^{-2}(0)$ and $\lambda^{-2}(0)_{BCS}$ as function of $T_c$; the *inset* shows the variation in $\Delta(0)$ as function of $T_c$. (d) Temperature variation of $(\lambda(T)/\lambda(0))^{-2}$ for the NbN film with $T_c = 2.27$K; the green solid line is a fit to the $T^2$ dependence of $\lambda^{-2}(T)$ at low temperature (T≤0.65K) and the red line shows the T dependence at higher temperature; the *inset* shows an expanded view of the $T^2$ dependence at low temperatures.

## Section 1

**Description of the in-situ, low temperature scanning tunneling microscope**

The temperature evolution of the tunneling DOS is measured using a home built high vacuum, low temperature, scanning tunneling microscope (STM) on epitaxial NbN films grown *in-situ* through reactive DC magnetron sputtering on (100) oriented single-crystalline MgO substrates.

Figure 1 shows schematics of the scanning tunneling microscope (STM). The substrate is mounted on a molybdenum sample holder (Figure 2), which is inserted in the deposition chamber using a horizontal manipulator with the substrate facing the sputtering gun. The substrate is heated through radiative heating using a molybdenum heater, located below the sample holder. The temperature is accurately controlled using a thermocouple sensor fitted inside the horizontal manipulator.

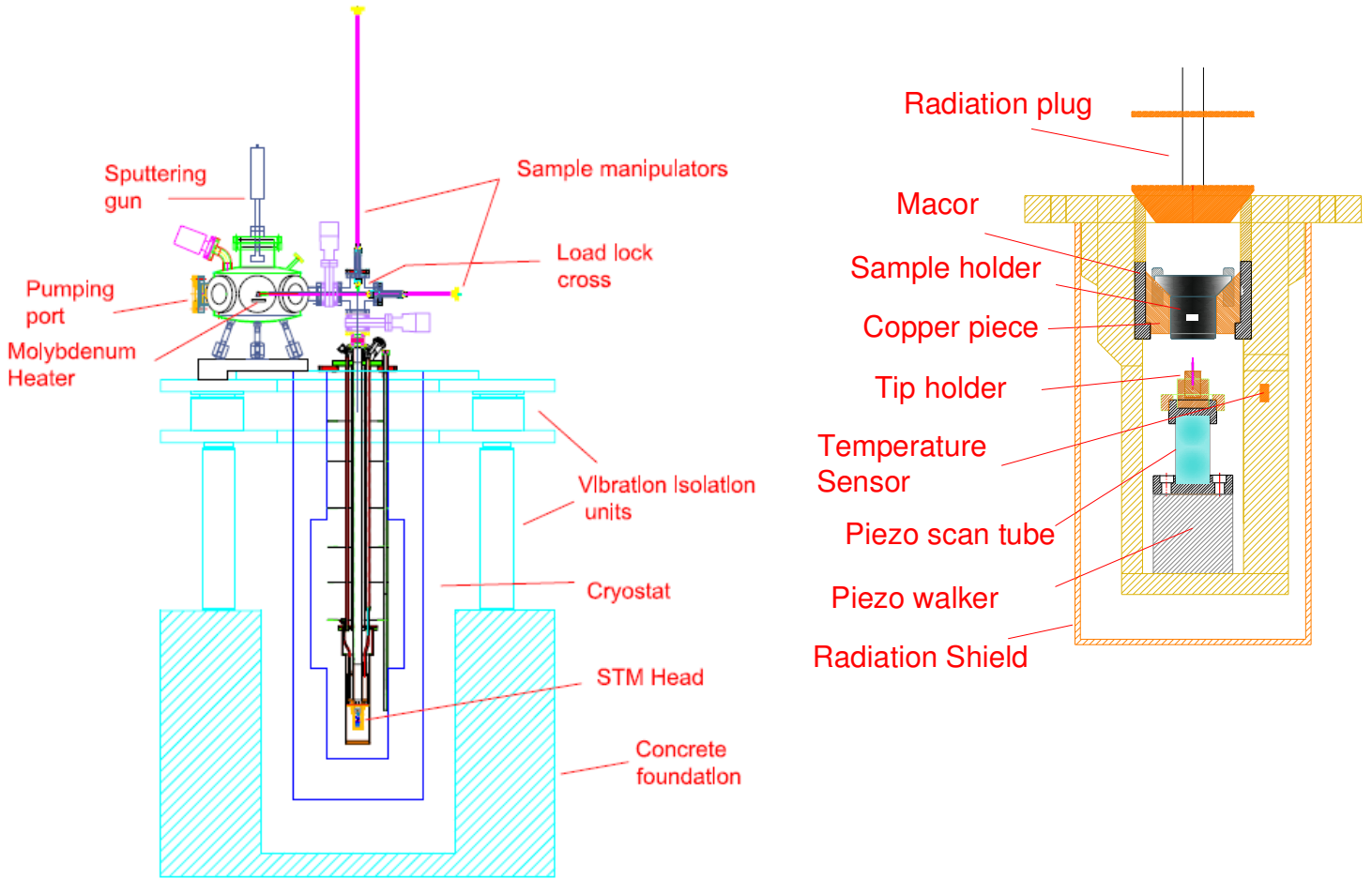

**Figure 1.** Schematic diagram of the low-temperature STM (left) and the expanded view of the STM head (right).

After the deposition, the sample is brought to a load-lock cross and transferred from the horizontal to a vertical manipulator which guides the sample holder directly into the STM head attached below the $^4$He pot inside the cryostat. To avoid any radiation falling on the sample holder, after loading the sample, a radiation plug is placed just above the STM head. Temperature uniformity is further ensured by placing the STM head inside a copper can. The entire cryostat assembly is isolated from building vibrations through several levels of vibration isolation.

Since NbN films have to be grown on insulating MgO substrates, one tricky part of these experiments is to ensure electrical contact between the film and the sample holder which acts as the sample electrode. To overcome this problem, we first deposit, *ex-situ,* strips of stoichiometric NbN on two sides of a 4mm by 4mm substrate (fig. 2). On loading this substrate on sample holder, the strips come in contact with the cap. The actual NbN film on which STM measurements are done, deposited *in-situ* through the hole in cap, is in contact with the strips. It is important to note that the region of the film on which the actual measurement is done is directly deposited on the bare MgO substrate and about 1.2mm away from contact pads on either side, thereby ruling out the possibility of any proximity effect from the contact pads.

We use RHK SPM 1000 controller as the control unit for the STM. For STS measurements the bias voltage is modulated with a fixed frequency (289 Hz) of small amplitude (100 µV) and the current signal is deconvoluted with a lock-in amplifier. Spatially resolved conductance curves are obtained by taking spectroscopy on each point of a defined grid.

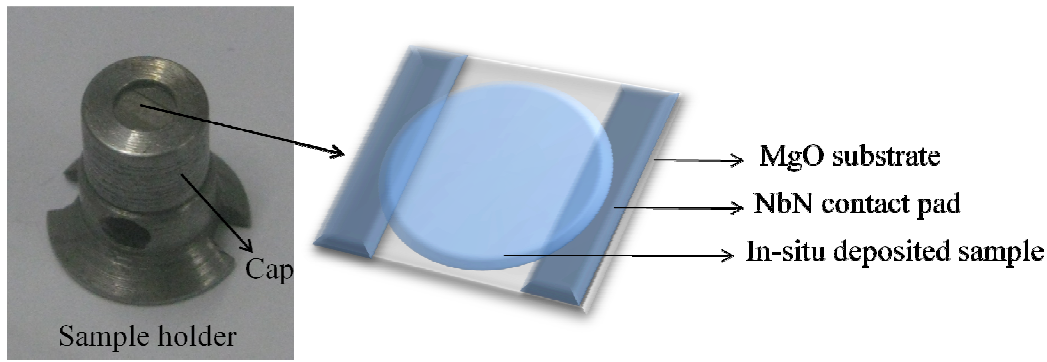

**Figure 2.** Sample holder with the substrate with contact pads (left) and schematic diagram of the substrate after *in-situ* sample deposition (right).

## Section 2.

## Evolution of the tunneling DOS at low temperature

We show here the evolution of the tunneling density of states as a function of disorder measured at low temperatures (<0.2$T_c$) on NbN/insulator/Ag planar tunnel junctions. The films on which the devices were fabricated have a range of $T_c$ varying from 14.9K down to 2.2K.

Figure 3 (a) shows a schematic of the device used which is similar to the one mentioned in Ref [17]. Using different wire configurations we can measure either the tunneling conductance of the NbN/insulator/Ag tunnel junction or the resistivity of the underlying NbN film. Figure 3 (b)-(f) show the tunneling density of states for a range of samples with different $T_c$. In the same plots we show the tunneling conductance calculated using the BCS density of states, $N(E) = \mathrm{Re}\left(\dfrac{E}{\sqrt{E^2 - \Delta^2}}\right)$, using the values of $\Delta$ shown in the panels. For the least disordered sample the tunneling data fits well with the BCS theory. However, as we go into the disordered regime, we observe a progressive suppression of the coherence peaks and corresponding appearance of states within the gap. We believe that the formation of these mid-gap states provides evidence for the emergence of a dissipative channel in the presence of strong disorder.

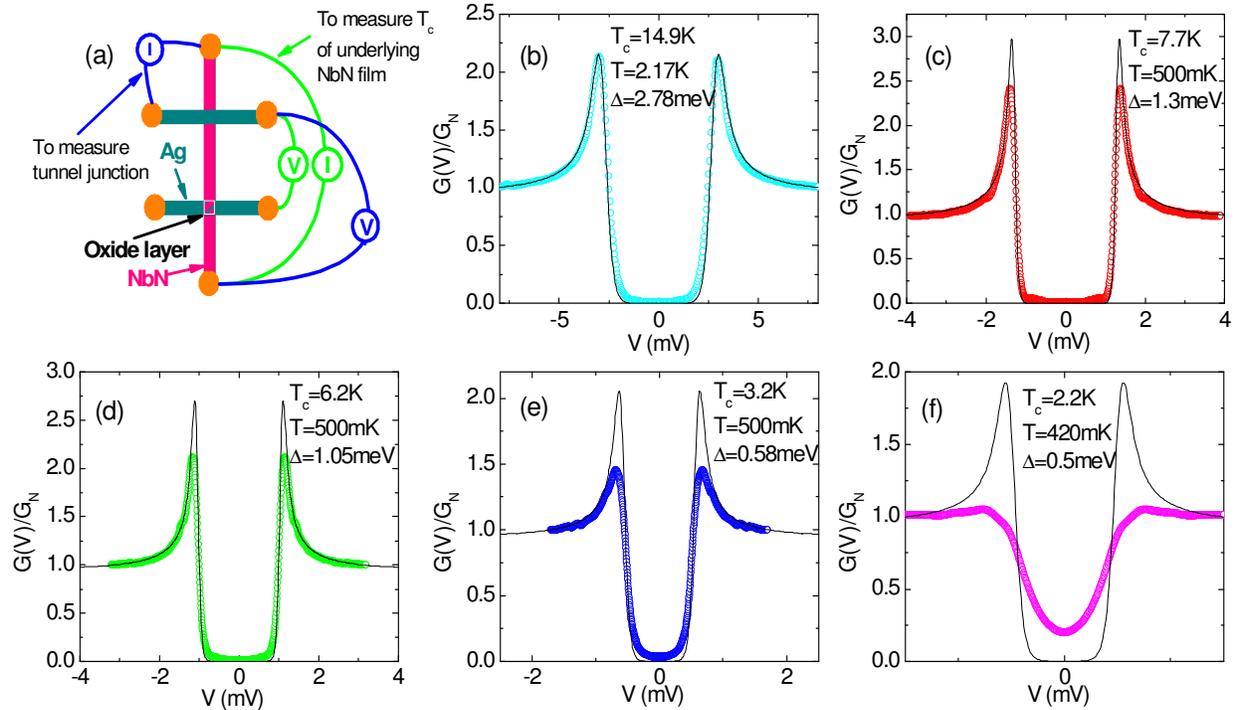

**Figure 3.** (a) Schematic drawing of the NbN/insulator/Ag planar tunnel junctions. Blue and green wire configurations correspond to tunneling I-V characteristic and resistivity measurements respectively. (b)-(f) Tunneling spectra for NbN samples with different $T_c$. In each case the solid black line represents the best fits using BCS density of states.

**Section 3:**

The temperature evolution of the tunneling DOS shown in **Fig. 2(a)-(d)** are obtained from the spectra averaged over 32 points along a 150 nm line on each sample. In order to show the evolution of the spatial inhomogeneity of the superconducting energy gap Figure 4 shows the tunneling DOS (at the lowest temperature) along the line in the form of an intensity plot of the normalized conductance, $G(V)/G_N$ (where $G_N$ is the conductivity at high bias), as a function of position and bias voltage on four samples with different $T_c$.

The line-scans reveal that with increase in disorder the superconducting state becomes progressively inhomogeneous. For the most disordered sample, where the zero bias conductance in the pseudogap state (at 2.65K) shows large spatial fluctuations, the line-scan suggests the formation of superconducting domains with size (5-15 nm) comparable to the Ginzburg-Landau coherence length estimated from the upper critical field[§].

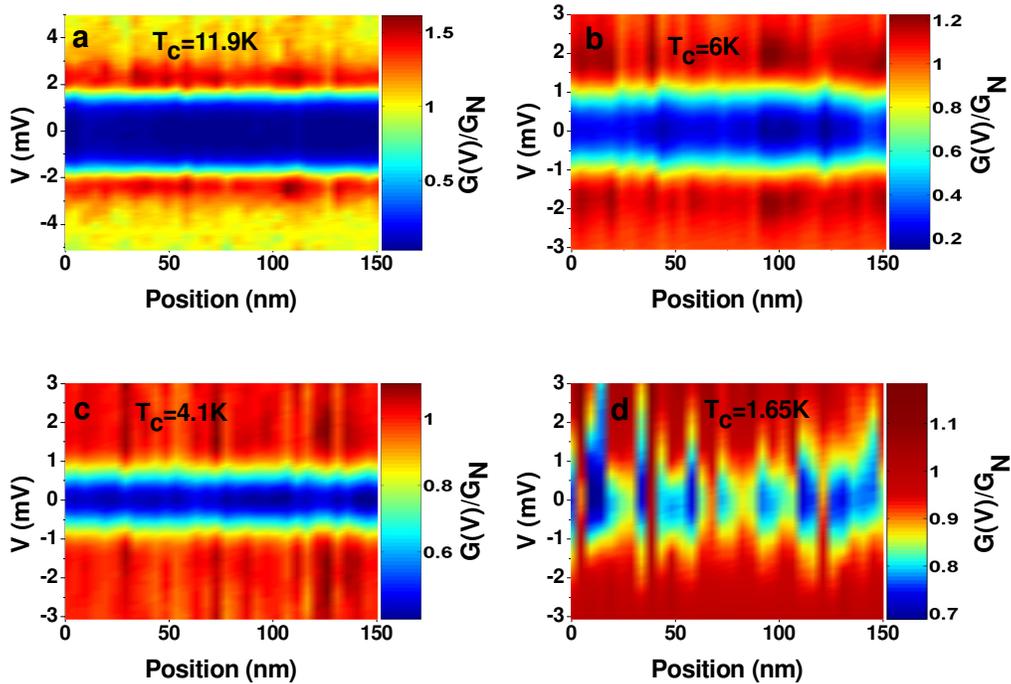

**Figure 4.** Intensity plot showing normalized tunneling conductance spectra acquired along the line of length 150 for four samples with $T_c$ ~ 11.9K, 6K, 4.1K, 1.65K. Spectra are acquired at temperatures 3.17K, 3.1K, 3.2K, 2.65K respectively.

---

[§] M. Mondal et al., arXiv:1005.1628 (2010).